\documentclass{article}

\usepackage[final, nonatbib]{neurips_2021} 

\usepackage[utf8]{inputenc} 
\usepackage[T1]{fontenc}    
\usepackage{hyperref}       
\usepackage{url}            
\usepackage{booktabs}       
\usepackage{amsfonts}       
\usepackage{nicefrac}       
\usepackage{microtype}      
\usepackage{xcolor}         
\usepackage{multirow}
\usepackage{multicol}
\usepackage{bbm}

\newcommand*\samethanks[1][\value{footnote}]{\footnotemark[#1]}

\title{Unsupervised Approaches for Out-Of-Distribution Dermoscopic Lesion Detection}

\author{
  Max Torop \\
  Northeastern University\\
  \texttt{torop.m@northeastern.edu} \\
   \And
   Sandesh Ghimire \\
   Northeastern University \\
   \texttt{sandesh@ece.neu.edu} \\
   \And
   Wenqian Liu \\
   Northeastern University \\
   \texttt{liu.wenqi@northeastern.edu} \\
   \And
   Dana H. Brooks \\
   Northeastern University \\
   \texttt{brooks@ece.neu.edu} \\
   \And
   Octavia Camps \\
   Northeastern University \\
   \texttt{o.camps@northeastern.edu} \\ 
   \And
   Milind Rajadhyaksha  \\
   Memorial Sloan Kettering Cancer Center \\
   \texttt{RajadhyM@mskcc.org} \\
   \And 
    Jennifer Dy\thanks{shared last authorship} \\
   Northeastern University \\
   \texttt{jdy@ece.neu.edu} \\
   \And
   Kivanc Kose\samethanks \\
   Memorial Sloan Kettering Cancer Center \\
   \texttt{kosek@mskcc.org} \\
}

\begin{document}

\maketitle

\begin{abstract}
There are limited works showing the efficacy of unsupervised Out-of-Distribution (OOD) methods on complex medical data. Here, we present preliminary findings of our unsupervised OOD detection algorithm, SimCLR-LOF, as well as a recent state of the art approach (SSD), applied on medical images. SimCLR-LOF learns semantically meaningful features using SimCLR and uses LOF for scoring if a test sample is OOD. We evaluated on the multi-source International Skin Imaging Collaboration (ISIC) 2019 dataset, and show results that are competitive with SSD as well as with recent \emph{supervised} approaches applied on the same data. 

\end{abstract}

\section{Introduction}

Although neural networks provide state of the art performance for image classification problems, including those in medical imaging \cite{StateArtMedImg3, StateArtMedImg2, StateArtMedImg1}, their success is mostly limited to problems with predefined classes that are supported by a large volume of labelled training data. However, in many settings, there are a significant number of diagnostic classes that occur infrequently, making it difficult to assemble fully labeled training data sets. Instead the data from these rare classes is frequently ignored while curating datasets \cite{GoogleHealthDerm}. These samples are referred to as Out-of-Distribution (OOD). Samples from classes that were included during training are referred to as In-Distribution (ID). It is a well known issue that models frequently provide over-confident incorrect predictions (i.e., high soft-max probability for an output class) when presented with OOD inputs \cite{GalThesis}. Therefore, automated detection of the OOD samples during/before model evaluation (e.g. for further manual inspection) has been a critical issue in adoption of machine learning models for clinical use. 

OOD detection algorithms for dermatological imaging, by and large, have been supervised learning methods that rely on laborious and costly labeling \cite{Combalia, ISIC2019WINNER, GRAM_SKIN}. A basic approach for OOD detection, which has been used for dermoscopy data, is training a model for $N+1$ classes where $N$ represents the number of known classes (ID classes) and any input data sample classified as OOD falls into the $N+1^{st}$ class \cite{ISIC2019WINNER}. However, in general, it is not realistic to have access to extensive amounts and types of OOD samples during training time, therefore these models do not generalize well to real clinical scenarios. Alternatively, some methods have been developed to utilize classification models that were trained only on ID classes, without needing OOD class labels. Pacheco et al. \cite{GRAM_SKIN} compare the similarity of the intermediate level features of a new test input sample to the training samples. Combalia et al. \cite{Combalia} quantify the uncertainty of the $N$-class classification model in predicting the diagnostic label of the test input via scoring the output softmax probabilities (Entropy, Variance, Bhattacharyya Distance) while using Monte-Carlo Dropout \cite{MCDROP} and/or Test Time Augmentation \cite{TTA} to get more robust uncertainty estimates. However, supervised models are also limited as they require labeled ID training data, which can itself be laborious to obtain. 

Unlike labelled data, unlabelled data is much easier to obtain. In this respect, there is a big need for OOD detection methods that can utilize unlabelled data. Unsupervised approaches, such as \cite{sehwag2021ssd, CSI2}, eliminate the cost of obtaining labels by leveraging large amounts of unlabeled data. In \cite{sehwag2021ssd}, Sehwag et al propose self-supervised outlier detection (SSD), that uses SimCLR \cite{SimCLR}, a contrastive feature learning algorithm, to learn features from the ID training set. After training, features are extracted from the ID train set and used to estimate the ID sample mean $\mu$ and covariance $\Sigma$. During inference, any input test point is scored against the ID train sample statistics ($\mu$ and $\Sigma$) using the Mahalanobis distance. SSD has shown state of the art results for a variety of natural image datasets. We use a similar approach, first learning features and then scoring them. 

In this work we use SimCLR \cite{SimCLR} to learn feature representations, and the Local Outlier Factor (LOF) \cite{LOF} method to provide ID scores for test points. SimCLR \cite{SimCLR}, trains the parameters $\theta$ of a network $f_\theta$, by attracting the feature representations of images $x$ and their (class preserving) augmented version $\hat{x}$ together while pushing features from other images away (preventing collapse to a trivial solution). The core idea is that features will be semantically meaningful if we learn them in such a way that they are invariant to these class preserving augmentations. Examples of non class preserving augmentations for skin would be random cropping an area so small that it does not contain any lesion, or overly blurring the lesion. The authors also show that contrasting features after they have been processed by an additional nonlinear projection head (a 2-3 layer ANN) leads to improved performance. We use the NT-Xent loss with cosine similarity for training in a contrastive manner.

Our scoring function; LOF \cite{LOF}, is a lazy learner, which takes in a dataset of ID features, a distance measure, and a neighborhood size $K$. The LOF score for a test feature vector is calculated by taking the ratio of the relative reachability of the test feature to the reachability of its K-nearest neighbors in the ID train set. Here the reachability of a feature vector is defined as one over its distance to its $K^{th}$ nearest neighbor. Intuitively the test point should be considered more OOD if the LOF score is high, as this means it is in a less reachable part of feature space compared to its K-nearest ID neighbors. 

In this paper we illustrate the performance of unsupervised OOD detection methods, our SimCLR-LOF as well as SSD, on the multi-source International Skin Imaging Collaboration (ISIC) 2019 challenge dataset \cite{MSK, BCN, HAM}, a commonly used benchmark for dermoscopic lesion classification. Our work has three main contributions. We provide early results indicating the potential for unsupervised  OOD detection in dermoscopy images. To our knowledge this is the first work using Local Outlier Factor (LOF) \cite{LOF} in conjunction with deep learning for OOD detection. Lastly, we find evidence for the difficulty gap between the HAM \cite{HAM} and BCN \cite{BCN} sources in ISIC 2019, knowledge which may benefit the community as ISIC 2019 is a popular benchmark.

\section{Methods and Experiments}
We train a resnet-18 based SimCLR model $f_\theta$ (discarding the classification head) on the ID training set, $X_{train} = \{x_i \}_{n=1}^N$ for feature learning. For scoring, we construct a LOF \footnote{used scikit-learn implementation} model using the features of the ID train set, $Z_{train} = f_\theta(X_{train}) = \{f_\theta(x_i)\}_{n=1}^N$, extracted by our resnet model. We then extract features for all ID test and OOD images, $Z_{IDtest} = f_\theta(X_{IDtest}), Z_{OOD} = f_\theta(X_{OOD})$, and feed them to our LOF model to get their outlier scores. Because SimCLR is trained using the cosine similarity, we use the inversely proportional cosine distance $d(z_1,z_2) = 1 - cossim(z_1, z_2)$ as the LOF distance measure.

\begin{table}
  \caption{SimCLR-LOF results vs SimCLR-Mahanolobis Distance (SSD) \cite{sehwag2021ssd} vs Monte-Carlo Dropout based method from \cite{Combalia} ({$^*$}the subset of the dataset that only contains NV samples is used as ID). The last column corresponds to CIFAR10 vs SVHN. }
    \vspace{-0.2cm}
  \label{Results}
  \centering
  \begin{tabular}{llllllll}
    \toprule
    SimCLR + LOF & ISIC 2019  & ISIC 2019$^*$ & HAM & HAM$^*$ & BCN & BCN$^*$ & CIFAR10\\
    \midrule
K = 10  & \bf{0.675} & 0.711 & 0.749 & 0.798 & \bf{0.642} & 0.664 & 0.841 \\
K = 50  & 0.668 & 0.749 & 0.774 & 0.884 & 0.602 & 0.646 & 0.894\\
K = 100 & 0.655 & 0.734 & 0.779 & 0.895 & 0.587 & 0.615 & 0.922\\
K = 200 & 0.648 & 0.739  & 0.770  & 0.887 & 0.567 & 0.604  & 0.941\\
K = 300 & 0.646 & 0.752 & 0.760  & 0.873 & 0.562 & 0.615  & 0.945 \\
\midrule
SSD  & 0.600  & \bf{0.800} & 0.789 & \bf{0.898}& 0.560 & \bf{0.721} & \bf{0.991} \\ 
\midrule
Supervised \cite{Combalia}  & - & - & \bf{0.800} & - & - & - & - \\ 
    \bottomrule\\
  \end{tabular}
  \vspace{-0.8cm}
\end{table} 

Our dataset of interest; the ISIC 2019 challenge dataset, lends itself to OOD detection as it has severe class imbalance. It consists of 25,531 images corresponding to 13,931 different lesions coming from three sources; Vienna (HAM) \cite{HAM}, Barcelona (BCN) \cite{BCN}, and New York (MSK) \cite{MSK}.  The BCN dataset is generally considered as the most difficult \cite{ISIC2019WINNER}; with hair obscuring lesional areas and/or images having black pinhole backgrounds of varying sizes. The dataset contains images from 8 lesion classes; Dermatofibroma (DF), Vascular lesion (VASC), Melanoma (MEL), Melanocytic nevus (NV), Basal cell carcinoma (BCC), Actinic keratosis (AK), Benign keratosis (BKL), and Squamous cell carcinoma (SCC). The dataset is severely imbalanced due to the challenge in finding images from rare lesion classes. As in \cite{Combalia}, we choose to treat the least common sample types ($~1\%$ each), DF and VASC as our OOD classes, and experiment with using either the remaining 6 classes, or the most frequent class, NV, as ID. We divide the ISIC 2019 dataset into approximately 80/5/15 percent, stratified train/validation/test splits, while ensuring that there is no leakage between splits.

As shown in Table \ref{Results} we evaluated SimCLR-LOF and compared against SSD \cite{sehwag2021ssd} and the supervised baseline in \cite{Combalia} on a number of subsets of the ISIC 2019 dataset. Similar to \cite{Combalia, sehwag2021ssd}, we used AUROC for ID vs OOD classes (with the LOF score as the varying threshold) for performance evaluation. In our first set of experiments (Table \ref{Results}-Column 1), we evaluate SimCLR-LOF using \emph{all} of the 6 ID classes from the train and test sets as well as all OOD images.

Columns 2-6 of Table~\ref{Results} use the same feature extractor (Trained on all ID training data from ISIC 2019), but, we fit and evaluate the LOF model on subsets of ID samples extracted features. During the scoring and evaluation phase of column 2 we restrict our consideration of the ID features to only NV. In column 3 LOF is only fit to, and evaluated on, image features from the HAM portion of the dataset. This restriction is applied to both the ID and OOD images we consider. Column 4 is the same as column 3 but only NV features from HAM are considered as ID. Columns 5 and 6 are the same as 3 and 4, but using BCN instead of HAM. We note that for all experiments, we maintain our OOD set as DF and VASC. In column 7, we show results for CIFAR10 (ID) vs SVHN (OOD) as validation that our algorithm can work well on easier benchmark datasets. 

\begin{table}[h]
\vspace{-0.4cm}
    
  \caption{SimCLR-LOF HAM vs BCN feature detection results.}
  \label{Invariance}
  \centering
  \begin{tabular}{lllllllll}
    \toprule
    \bf{K} &40 & 50 & 60 & 70 & 100 & 200 & 250 & 300 \\ 
    \midrule
    HAM vs BCN 6  & 0.946  & 0.948 & 0.951 & 0.952 & 0.956 & 0.956 & 0.955 & 0.954  \\
    HAM vs BCN NV  & 0.955 & 0.958 & 0.959 & 0.960 & 0.961 & 0.954 & 0.950 & 0.947 \\
    \bottomrule\\
    \vspace{-1.5cm}
  \end{tabular}
\end{table}

\section{Discussion and Conclusion}

For the experiments presented in columns 2-6 of Table \ref{Results} we use the feature extractor trained on all ID train data because we interestingly found that this provides significantly better results. Specifically, we found that the inclusion of BCN in SimCLR training leads to more robust HAM features, and the inclusion of the other 5 ID classes leads to more robust NV features. We experiment with different neighborhood sizes K for LOF, where  smaller neighborhoods may be noisier and denser while larger neighborhoods may be smoother and more dispersed. It can be seen that there are different optimal values of K for different datasets. 

Table \ref{Invariance} shows that SimCLR is learning very different features for HAM and BCN images. For each row in Table \ref{Invariance}, we use the SimCLR feature extractor trained on all ID train data. In the first row we evaluate the AUROC (at different K) for distinguishing between HAM and BCN where we fit LOF to ID train HAM and evaluate using ID test HAM as ID and ID test BCN as OOD. LOF is able to pick up the difference between HAM and BCN features with high AUROC. This suggests that SimCLR is learning very different features for HAM and BCN data, which is not optimal as we would prefer for our features to be invariant to the dataset, and only encode semantically meaningful parts of the images. To further assess this conclusion we repeat the experiment but using only the NV images from HAM and BCN, ensuring that this feature difference is not coming from the difference in class distributions between the two datasets. As can be seen in row 2, the AUROC is still high indicating that features learned for HAM NV are quite different then those learned for BCN NV. 

It is noteworthy that for the unsupervised methods the highest HAM AUROCs 0.779/0.789, are much better then the highest BCN AUROCs 0.642/0.560, due to the difficulty of the BCN dataset. It is also noteworthy that, when restricting HAM LOF fitting to NV from ID, the best AUROC rises from 0.779/0.789 to 0.895/0.898. This may be due to the fact that the majority of the HAM train set is NV (67 \%), resulting in robust NV HAM features from SimCLR as well as larger and more robust NV clusters for the LOF fitting. This effect may not be seen in BCN because BCNs class distribution for NV (34 \%) is not as biased.

Our results for HAM can be compared to the results in \cite{Combalia} as they run a very similar experiment on similar partitions. We were able to achieve comparable results to those in \cite{Combalia}, without using any labels. However, as unsupervised approaches, SimCLR-LOF and SSD can easily be scaled to massive datasets - with the potential for major performance gains. 

\section*{Societal Impact}

Automating aspects of diagnosis with deep learning (DL) classifiers has the potential to save doctors time, allowing for the quick diagnosis of more patients. The development of OOD detection algorithms for dermoscopy data is important for the clinical usage of DL classifiers. While this paper shows the potential for such algorithms on dermoscopy data, it is preliminary and much work has to be done before these algorithms can be used in practice. An important issue with current dermoscopy classifiers is that popular datasets, such as ISIC 2019, consist mainly of light skinned patients. Before OOD detection algorithms or DL classifiers are used in practice it is essential that these methods are developed for and evaluated on dermoscopy images covering a more diverse range of skin color. 

\section*{Acknowledgments}
This project was supported by  NSF grants IIS-1814631 and  CPS-2038493, NIH grants R01CA199673, R01CA240771 from NCI and in part by MSKCC’s Cancer Center core support NIH grant P30CA008748 from NCI.



\end{document}